\documentclass[conference]{IEEEtran}
\IEEEoverridecommandlockouts
\usepackage{cite}
\usepackage{amsmath,amssymb,amsfonts}
\usepackage{algorithm}
\usepackage{algorithmic}
\usepackage{graphicx}
\usepackage{multirow}
\usepackage{booktabs}
\usepackage{textcomp}
\usepackage{xcolor}
\def\BibTeX{{\rm B\kern-.05em{\sc i\kern-.025em b}\kern-.08em
    T\kern-.1667em\lower.7ex\hbox{E}\kern-.125emX}}
\begin{document}

\title{{\fontsize{22}{26}\selectfont SecRTL-Bench: Benchmarking LLMs on IP-Level RTL Vulnerability Detection, Localization, and Repair\\}}

\author{%
\IEEEauthorblockN{\textbf{Qian Peng, Zhenwei Guo, Sibo Ma, Li Kuang\textsuperscript{*}}}
\IEEEauthorblockA{\normalfont School of Computer Science and Engineering, Central South University, Changsha, China}
}

\maketitle

\begin{abstract}
Large language models are increasingly applied to RTL security analysis, yet existing benchmarks often lack realistic design complexity, reproducible evidence of security violations, and unified evaluation across vulnerability detection, localization, and repair. To address these limitations, we present SecRTL-Bench, a unified RTL security benchmark constructed from real-world open-source IP-level designs through a multi-agent framework. The framework identifies security assets and protection mechanisms, generates source-grounded security contracts to guide vulnerability injection, and employs automatically calibrated verification environments and EDA simulation to validate actual security violations. Through execution feedback and independent repair, it produces reproducible vulnerability instances with functional and security verification packages, ground-truth annotations, and security violation evidence. SecRTL-Bench comprises 413 vulnerability instances spanning 31 hardware CWE weaknesses. We evaluate nine advanced LLMs using a composite scoring mechanism that jointly considers CWE identification accuracy, line-level localization accuracy, and the functional correctness and security of repair patches. The best-performing model achieves an overall score of only 59.83/100. Further analysis reveals that security verification pass rates consistently lag behind functional pass rates, highlighting the difficulty of generating reliable patches that satisfy both functional and security requirements. Our data and code are available at https://github.com/QianPeng369/SecRTL-Bench.
\end{abstract}


\section{Introduction}
Hardware security vulnerabilities discovered late in the chip design process or after fabrication often require design modifications, re-verification, or even re-fabrication, resulting in substantial remediation costs \cite{bib1}. Therefore, shifting security analysis to the RTL stage  is essential for early vulnerability detection, accurate localization, and effective repair \cite{bib2}.

In recent years, Large Language Models (LLMs) have increasingly been applied to Register-Transfer Level (RTL) vulnerability detection, localization, and repair, making high-quality benchmarks essential for objectively evaluating their capabilities. Existing RTL security benchmarks mainly contain two types of vulnerability samples. The first constructs vulnerability instances around specific hardware common weakness enumerations (CWEs) using relatively independent RTL modules or code snippets, typically providing vulnerable code with corresponding CWE labels or vulnerability locations \cite{bib3,bib4,bib6,bib7,bib10,bib21}. The second focuses on specific hardware security scenarios, such as processor privilege control, sensitive resource access, and hardware Trojans \cite{bib5,bib8,bib9}. Together, these studies support the development and evaluation of security analysis methods, covering both code-level weaknesses and hardware security scenarios.

\begin{figure}[t]
\centering
\includegraphics[width=\columnwidth]{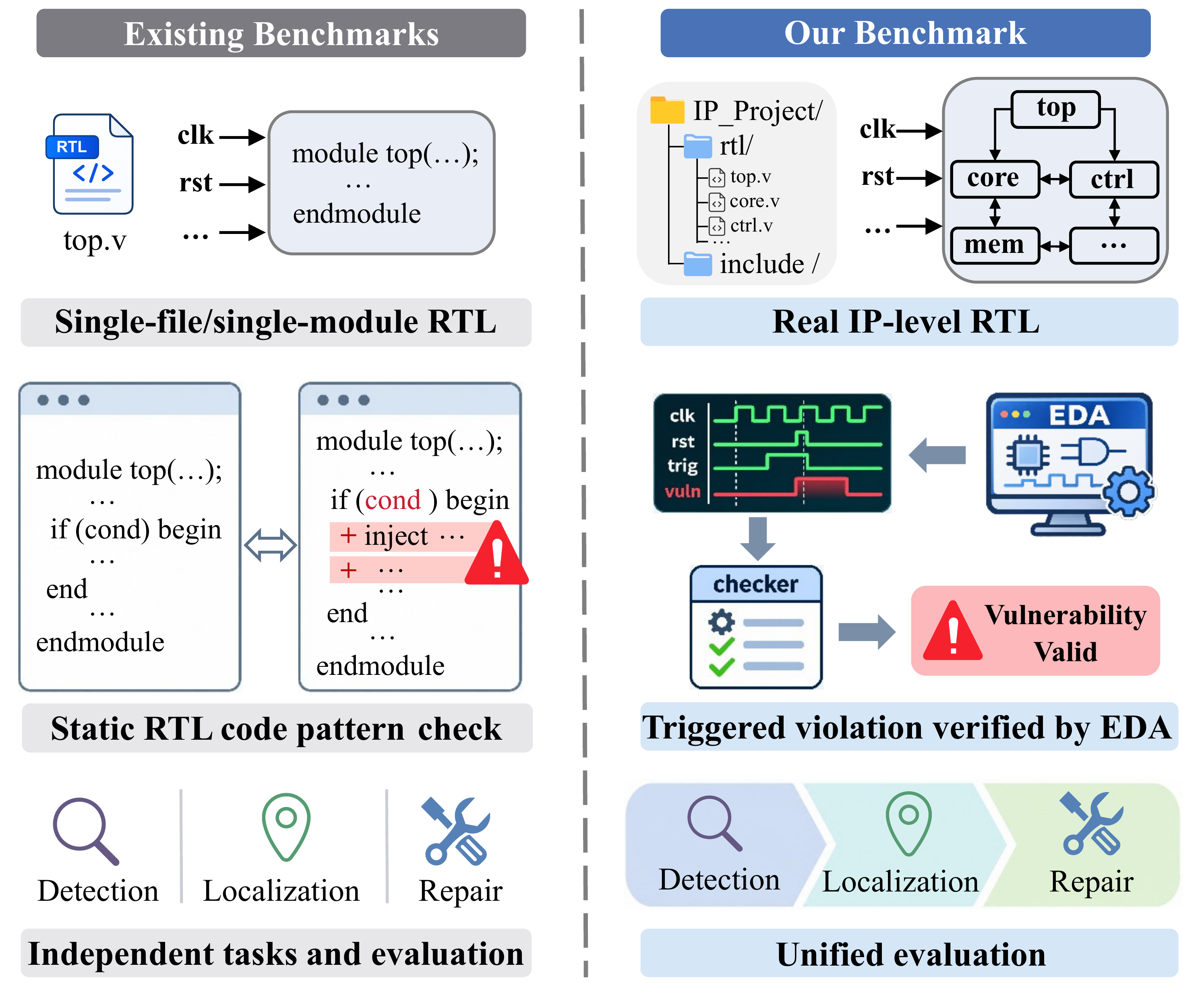}
\caption{Comparison of existing RTL security benchmarks and our benchmark}
\label{fig1}
\end{figure}

However, existing RTL security benchmarks still have limitations in data complexity, vulnerability validity, and task relevance, as shown in Fig.~\ref{fig1}. \textbf{First}, most benchmarks rely on simplified RTL modules, failing to capture the module dependencies and complex control logic of real-world SoC/IP designs. \textbf{Second}, vulnerability injection is often validated through static code pattern matching or analysis of logical differences before and after injection, without verifying whether the injected vulnerability can actually be triggered by specific stimuli. \textbf{Furthermore}, existing benchmarks typically evaluate vulnerability detection, localization, and repair as independent tasks, overlooking their inherent interdependence and sequential nature. The lack of a unified evaluation framework based on the same vulnerability instances makes it difficult to comprehensively assess the end-to-end capabilities of LLMs from vulnerability discovery to repair. Therefore, this paper aims to construct an RTL security benchmark based on real-world IP-level designs with reproducible evidence of security violations, providing a unified evaluation foundation for vulnerability detection, localization, and repair.

To achieve the aforementioned benchmark construction goals,  two key challenges remain:

\textbf{Challenge 1: Difficulty in identifying vulnerability injection locations in multi-file RTL designs.} Real-world RTL projects involve multiple source files, complex module dependencies, and control logic. Their security properties are often implemented by multiple interdependent logic components, making it difficult to accurately map abstract CWE weaknesses to specific RTL code locations for vulnerability injection.

\textbf{Challenge 2: Inability to verify the effectiveness of injected vulnerabilities through static code analysis.} Triggering RTL vulnerabilities depends on specific input conditions, state transitions, and timing relationships. Even when vulnerability injection modifies the original security logic, observable security violations may not occur. Consequently, RTL modifications and conventional functional simulation alone are insufficient to reliably validate the effectiveness of vulnerability injection.

To address these challenges, this paper proposes a multi-agent collaborative framework for automatically constructing security benchmarks from real-world IP-level designs. The framework uses security contracts to guide vulnerability construction and validation, integrating multi-agent collaboration with EDA execution feedback to coordinate security asset identification, vulnerability injection, dynamic validation, and independent repair.  \textbf{For Challenge 1}, we develop a security asset-centric mechanism for identifying vulnerability injection targets. By jointly analyzing RTL module hierarchies, signal dependencies, and security protection logic, the mechanism assesses the applicability of CWE weaknesses to specific designs in terms of controllability, reachability, and observability. It then generates security contracts backed by source code evidence to provide semantic guidance for vulnerability construction and injection locations. \textbf{For Challenge 2}, we further design a dynamic verification mechanism based on actual security violations. Guided by security contracts, the mechanism automatically generates and calibrates vulnerability-triggering stimuli and security checkers, reproduces security violations through EDA simulation, and compares execution results before and after injection to determine whether the injected vulnerability violates the protection constraints of the target security assets. This produces reproducible and replayable evidence of vulnerability validity. \textbf{Finally}, independent repair and regression validation are performed to construct RTL security vulnerability samples with complete validation evidence, providing a unified evaluation foundation for vulnerability detection, localization, and repair.

Using the automated construction framework, we build SecRTL-Bench, where each vulnerability instance is derived from a real-world IP-level design and includes vulnerable RTL code, build configurations, functional and security verification packages, CWE labels and vulnerability locations, and reproducible evidence of security violations. To evaluate models' overall performance across RTL vulnerability detection, localization, and repair, we design a composite scoring mechanism that accounts for both diagnostic quality and repair effectiveness by jointly evaluating CWE identification accuracy, line-level localization accuracy, and the functional correctness and security of generated repair patches. Our evaluation of nine advanced LLMs shows that the best-performing model achieves an overall score of only 59.83, highlighting the challenges of real-world IP-level RTL security analysis. Further analysis reveals that security verification pass rates are consistently lower than functional pass rates, indicating that generating effective patches satisfying both functional and security requirements remains a major challenge for current LLMs.

The contributions of this paper are as follows:

\begin{itemize}
\item We propose a multi-agent collaborative framework for automatically constructing security benchmarks from real-world IP-level designs. The framework integrates security contract guidance, dynamic vulnerability validation, and execution feedback to automate the construction and validation of security vulnerability samples.
\item We construct SecRTL-Bench, an RTL security benchmark based on real-world open-source IP designs, comprising 413 vulnerability samples spanning 31 hardware CWE weaknesses. It provides unified evaluation data and ground truth for vulnerability detection, localization, and repair.
\item Using SecRTL-Bench, we comprehensively evaluate nine advanced LLMs and report our findings on their capability differences and vulnerability repair reliability in real-world RTL security analysis.
\end{itemize}

\section{The SecRTL-Bench Benchmark}
\begin{figure*}[t]
\centering
\includegraphics[width=0.96\textwidth]{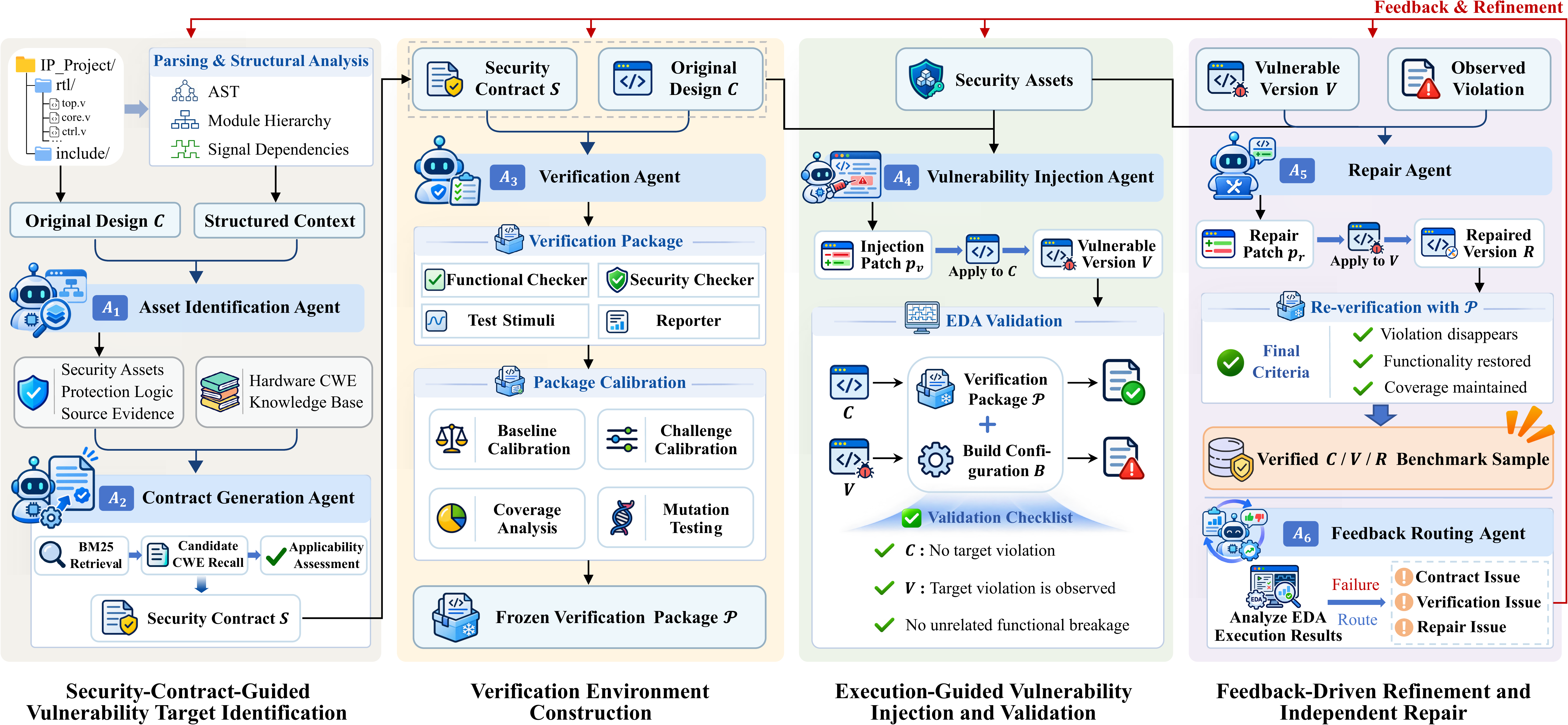}
\caption{Overview of SecRTL-Bench construction framework}
\label{fig2}
\end{figure*}

This section presents the automated construction framework of SecRTL-Bench, as illustrated in Fig.~\ref{fig2}. The framework takes RTL designs from real-world open-source SoC/IP projects and a hardware CWE knowledge base as inputs. Six specialized agents sequentially perform vulnerability target identification, verification environment construction, vulnerability injection, validity verification, and independent repair. A unified controller coordinates these stages and drives iterative refinement based on actual EDA execution results, ultimately producing RTL vulnerability benchmark samples with executable security evidence.

\subsection{Security-Contract-Guided \small Vulnerability Target Identification}
To identify security-relevant vulnerability injection targets in real-world RTL projects, we first employ hardware analysis tools (e.g., Yosys and slang) to parse and elaborate IP-level RTL, extracting abstract syntax tree (AST), module hierarchies, and signal dependencies. Based on this structural information and RTL source code, the asset identification agent ($A_1$) analyzes the functional semantics and access/update relationships of critical registers, control states, and interface signals to identify security assets and protection mechanisms, linking them to specific signals and source locations.

The contract generation agent ($A_2$) then constructs queries from the identified assets, protection mechanisms, and functional semantics, using BM25 to retrieve candidate CWEs by matching their descriptions, mitigations, and examples. It evaluates each candidate's applicability based on controllability, reachability, observability, tool support, and injectability. For applicable candidates, $A_2$ generates a security contract $\mathcal{S}$ specifying protected assets, normal behavior, security constraints, and triggering conditions, linked to corresponding RTL signals and protection logic. This source-grounded mapping supports traceability from CWE weaknesses to design-specific security requirements and injection targets.

\subsection{Verification Environment Construction}
To prevent verification rules from being tailored to injected vulnerabilities, we construct and calibrate the verification environment before vulnerability injection. Based on the original RTL design $C$ and security contract $\mathcal{S}$, the verification agent ($A_3$) generates a verification package $\mathcal{P}$ containing functional and security checkers, test stimuli, and result loggers to verify functionality, detect security violations, and record execution results.

To address incorrect expectations, insufficient stimuli, and ineffective checkers, we adopt a two-stage calibration mechanism. During Baseline Calibration, EDA simulation verifies that $C$ satisfies functional and security requirements under tested scenarios. During Challenge Calibration, controlled security-related modifications test whether $\mathcal{P}$ can detect expected violations. Functional mutation testing and code coverage analysis further assess verification sensitivity and coverage. Once calibrated, $\mathcal{P}$ is frozen and reused unchanged during vulnerability injection and repair, preventing verification rules from being tailored to subsequent patches.

\subsection{Execution-Guided Vulnerability Injection and Validation}
The vulnerability injection agent ($A_4$) generates a patch $p_v$ guided by the security contract $\mathcal{S}$, security assets, and protection logic, which is applied to the original design $C$ to obtain the vulnerable version $V$.

To validate the effectiveness of injected vulnerabilities, we develop an EDA-driven dynamic security violation validation mechanism. Both $C$ and $V$ are simulated under the same build configuration and frozen verification package $\mathcal{P}$, using targeted stimuli and security checkers to assess violations of $\mathcal{S}$. A candidate is accepted only if $C$ exhibits no target violation, $V$ triggers the expected violation, and no unrelated functional failures occur. We retain injection patches, source locations, and execution evidence, enabling security violations to be traced and replayed under identical conditions.

\subsection{Feedback-Driven Refinement and Independent Repair}
Vulnerability construction may fail due to inapplicable security contracts, verification environment errors, or unsuccessful injection. To address these failures, the feedback agent ($A_6$) analyzes EDA execution results to identify their causes, while the Controller routes tasks to the corresponding stages for revision. Changes to upstream contracts or verification packages are versioned, and affected downstream results are revalidated to maintain consistency.

Once $V$ passes validation, the repair agent ($A_5$) identifies the vulnerability's root cause using the vulnerable RTL, protection requirements in $\mathcal{S}$, and observed security violations. It then generates a repair patch $p_r$ to produce $R$. To prevent answer leakage, $A_5$ cannot access the original clean design $C$ or injection patch $p_v$. The repaired version $R$ is verified using the frozen package $\mathcal{P}$, requiring elimination of the target security violation, restoration of normal functionality, and preservation of necessary test coverage. The framework thus produces complete $C$/$V$/$R$ samples with corresponding verification evidence.

\begin{figure}[t]
\centering
\includegraphics[width=0.85\columnwidth]{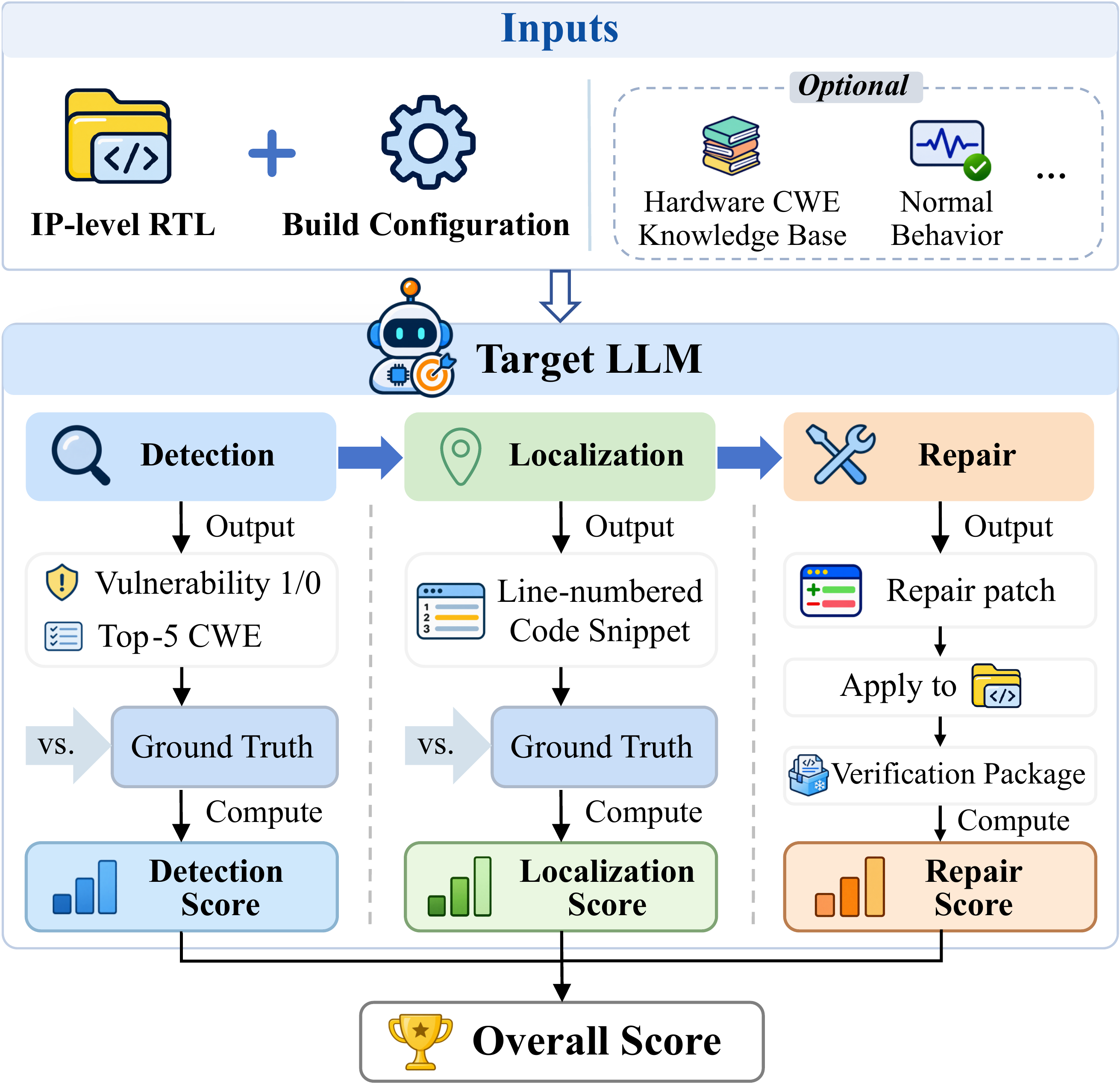}
\caption{SecRTL-Bench evaluation architecture}
\label{fig3}
\vspace*{-\baselineskip}
\end{figure}

\subsection{Evaluation Architecture}
As shown in Fig. 3, SecRTL-Bench takes real-world IP-level RTL designs and build configurations $B$ as inputs, with optional CWE knowledge and normal functionality descriptions. The target LLM sequentially performs vulnerability detection, localization, and repair. For detection, the model predicts vulnerability presence and Top-5 CWE candidates, which are evaluated against ground truth. For localization, it identifies vulnerable code snippets with file paths and line numbers, evaluated by line-level matching. For repair, the generated patch is applied to the vulnerable RTL and assessed for functional correctness and security using the frozen verification package $\mathcal{P}$. Each stage uses the preceding predictions without ground-truth correction. Finally, the three task results are combined into an overall score.

Since vulnerability detection, localization, and repair are interdependent sequential tasks, evaluating them independently cannot capture their combined impact. Therefore, we propose a composite scoring mechanism that considers both diagnostic quality and repair effectiveness. Specifically, for the $i$-th vulnerability instance, let $d_i,\ell_i\in[0,1]$ denote CWE MRR@5 and line-level localization F1, respectively. The diagnostic quality score $q_i$ is calculated as follows:
\begin{equation}
q_i=\alpha \cdot d_i+\beta\cdot\ell_i+\gamma\cdot\min(d_i,\ell_i)
\label{eq1}
\end{equation}
where $\alpha,\beta,\gamma\geq0$ and $\alpha+\beta+\gamma=1$. $\alpha$ and $\beta$ weight detection and localization, respectively, while $\gamma$ rewards strong performance in both. Let $f_i,s_i\in{0,1}$ denote functional preservation and security verification outcomes for the $i$-th instance. The final overall score is averaged across $N$ vulnerability instances:
\begin{equation}
\small
\text{\textit{Overall Score}}=
\frac{1}{N}\sum_{i=1}^{N}
\left(a q_i+b f_i+c s_i+\delta f_is_i+\eta q_if_is_i\right)
\label{eq2}
\end{equation}
where $a$, $b$, and $c$ weight diagnostic quality, functional preservation, and security verification, respectively. $\delta$ rewards joint functional and security correctness, while $\eta$ rewards accurate diagnosis with complete repair. These coefficients are non-negative and sum to 100, ensuring complete repairs score higher than incomplete ones while retaining each stage's contribution\cite{bib11,bib12}.

\begin{table}[b]
\vspace*{-\baselineskip}
\caption{\scriptsize Statistics of SecRTL-Bench. Mean instance size is computed across vulnerability instances.}
\label{tab:secrtl-bench-statistics}
\centering
\scriptsize
\renewcommand{\arraystretch}{1.12}
\setlength{\tabcolsep}{2.5pt}
\begin{tabular*}{\columnwidth}{@{\extracolsep{\fill}}lrrrrrr}
\hline
\multirow{2}{*}{\textbf{Source}} & \multicolumn{3}{c}{\textbf{Coverage}} & \multicolumn{3}{c}{\textbf{Mean instance size}} \\
\cline{2-4}\cline{5-7}
 & \textbf{IPs} & \textbf{Vulns.} & \textbf{CWEs} & \textbf{Files} & \textbf{Mods.} & \textbf{LoC (\textit{K})} \\
\hline
Caliptra & 17 & 28 & 11 & 14.79 & 10.71 & 2.79 \\
CV32E40S & 11 & 22 & 9 & 5.77 & 4.73 & 3.51 \\
CVA6 & 20 & 40 & 9 & 9.73 & 3.40 & 3.62 \\
FuseSoC & 41 & 71 & 9 & 5.35 & 6.44 & 1.52 \\
OpenCores & 93 & 112 & 16 & 2.89 & 5.21 & 0.68 \\
OpenTitan & 29 & 59 & 14 & 82.46 & 68.83 & 15.24 \\
PULP & 31 & 48 & 12 & 4.00 & 5.56 & 0.65 \\
XiangShan & 28 & 33 & 9 & 9.64 & 9.64 & 4.22 \\
\hline
\textbf{Total} & \textbf{270} & \textbf{413} & \textbf{31} & \textbf{16.97} & \textbf{15.08} & \textbf{3.76} \\
\hline
\end{tabular*}
\end{table}

\section{Evaluation}

\begin{table*}[t]
\vspace*{-\baselineskip}
\caption{\scriptsize Unified Evaluation Results on SecRTL-Bench. Metrics are reported in \%, except Overall Score (0--100). Best in bold, second-best underlined.}
\label{tab:unified-evaluation-results}
\centering
\tiny
\setlength{\tabcolsep}{3pt}
\renewcommand{\arraystretch}{1.12}
\setlength{\aboverulesep}{0.5pt}
\setlength{\belowrulesep}{0.5pt}
\resizebox{0.97\textwidth}{!}{%
\begin{tabular}{l*{14}{c}}
\hline
\multirow{2}{*}{\textbf{Models}} & \multicolumn{2}{c}{\textbf{Detection Rate}} & \multicolumn{2}{c}{\textbf{CWE MRR@5}} & \multicolumn{2}{c}{\textbf{Localization F1}} & \multicolumn{2}{c}{\textbf{Functional Pass}} & \multicolumn{2}{c}{\textbf{Security Pass}} & \multicolumn{2}{c}{\textbf{Repair Success}} & \multicolumn{2}{c}{\textbf{Overall Score}} \\
\cmidrule(lr){2-3}\cmidrule(lr){4-5}\cmidrule(lr){6-7}\cmidrule(lr){8-9}\cmidrule(lr){10-11}\cmidrule(lr){12-13}\cmidrule(lr){14-15}
 & E0 & E1 & E0 & E1 & E0 & E1 & E0 & E1 & E0 & E1 & E0 & E1 & E0 & E1 \\
\hline
\multicolumn{15}{c}{\textit{Closed-source Models}} \\
\hline
GPT-6.1-Sol & 67.07 & 69.25 & 14.58 & 36.28 & 35.03 & 36.12 & 58.84 & 63.44 & 54.72 & 61.02 & 54.48 & 60.53 & 37.74 & 47.66 \\
GPT-6-Astra & 61.17 & 68.45 & 6.08 & 29.31 & 22.42 & 22.95 & 47.33 & 50.00 & 40.29 & 43.45 & 40.05 & 42.96 & 26.30 & 34.49 \\
Claude-Sonnet-5.5 & \underline{82.73} & \underline{83.62} & 6.08 & 15.63 & 13.03 & 10.46 & 47.45 & 43.28 & 34.55 & 29.10 & 32.60 & 27.63 & 22.19 & 21.84 \\
Claude-Opus-5.5 & \textbf{83.50} & \textbf{90.93} & 3.54 & 14.87 & 10.88 & 4.59 & 44.66 & 41.67 & 31.80 & 28.19 & 29.85 & 25.74 & 19.80 & 19.85 \\
Gemini-4-Argon & 78.69 & 73.89 & \textbf{18.95} & \textbf{49.14} & \textbf{54.78} & \textbf{51.20} & \textbf{76.51} & \textbf{72.41} & \textbf{76.03} & \textbf{71.43} & \textbf{75.30} & \textbf{70.94} & \textbf{52.38} & \textbf{59.83} \\
\hline
\multicolumn{15}{c}{\textit{Open-source Models}} \\
\hline
Qwen3.5-35B-A3B & 42.93 & 70.54 & 0.25 & 14.32 & 2.70 & 2.06 & 16.67 & 25.14 & 5.81 & 11.35 & 5.56 & 10.54 & 4.85 & 10.29 \\
Qwen3.8-Flash & 43.80 & 64.27 & 0.43 & 22.92 & 16.78 & 20.50 & 32.85 & 47.64 & 27.98 & 39.45 & 27.74 & 38.96 & 17.10 & 30.50 \\
GLM-5.3-Flash & 32.76 & 38.83 & 1.87 & 13.80 & 12.27 & 11.37 & 23.47 & 25.53 & 20.54 & 19.68 & 19.80 & 19.68 & 12.88 & 15.94 \\
DeepSeek-V4.1-Flash & 72.64 & 75.18 & \underline{14.89} & \underline{39.06} & \underline{35.48} & \underline{37.39} & \underline{65.62} & \underline{67.88} & \underline{61.74} & \underline{64.23} & \underline{60.53} & \underline{63.02} & \underline{41.14} & \underline{50.52} \\
\hline
\end{tabular}%
}
\vspace*{-\baselineskip}
\end{table*}

\subsection{Benchmark Statistics}
We collect IP-level RTL designs from eight open-source hardware repositories: Caliptra\cite{bib13}, CV32E40S\cite{bib14}, CVA6\cite{bib15}, FuseSoC\cite{bib16}, OpenCores\cite{bib17}, OpenTitan\cite{bib18}, PULP\cite{bib19}, and XiangShan\cite{bib20}. After preprocessing and filtering, 707 original IP/configuration entries are retained, covering various hardware functions, including RISC-V processors, cryptographic operations, key management, memory and access control, communication interfaces, roots of trust (RoT), and direct memory access (DMA). SecRTL-Bench is then generated from these IPs using our automated multi-agent construction framework. As shown in Table~\ref{tab:secrtl-bench-statistics}, SecRTL-Bench comprises 270 IPs and 413 vulnerability instances spanning 31 hardware CWE weaknesses. Each instance contains an average of 16.97 RTL files, 15.08 modules, and 3.76K lines of code, with the largest instance containing 147 RTL files and 46.79K lines of code, reflecting the engineering complexity of real-world IP-level designs. Of the 413 vulnerability instances, 331 involve single-line, single-location modifications, while the remaining 82 involve modifications across multiple lines or logic locations.

\subsection{Experimental Setup}
This paper evaluates nine representative large language models, including five closed-source models: GPT-6.1-Sol, GPT-6-Astra, Claude-Sonnet-5.5, Claude-Opus-5.5, and Gemini-4-Argon; as well as four open-source models: Qwen3.5-35B-A3B, Qwen3.8-Flash, GLM-5.3-Flash, and DeepSeek-V4.1-Flash-1017. The EDA toolchain uses Icarus Verilog and Verilator for RTL compilation, simulation, and syntax checking, together with Yosys and its Slang frontend for RTL parsing, module hierarchy analysis, and signal dependency extraction. During benchmark construction, all agents are powered by GPT-6.1-Sol. All evaluated models follow the same vulnerability detection, localization, and repair workflow, with functional and security verification performed using the frozen verification package. We consider two input settings: E0 provides IP-level RTL source code, build configurations, and normal functionality descriptions, while E1 additionally provides a retrievable Hardware Design CWE-1194 View knowledge base without revealing the target vulnerability's CWE label. In addition, we report the vulnerability detection rate, evaluated on both 413 vulnerability instances and 270 original IP designs without injected vulnerabilities, mean reciprocal rank@5 of CWE candidates (MRR@5), line-level localization F1, functional verification pass rate, security verification pass rate, and repair success rate. For the overall score, we assign weights of $(0.6, 0.3, 0.1)$ to $(\alpha,\beta,\gamma)$ and $(23,15,24,11,27)$ to $(a,b,c,\delta,\eta)$, respectively.

\begin{figure}[t]
\centering
\includegraphics[width=0.9\columnwidth]{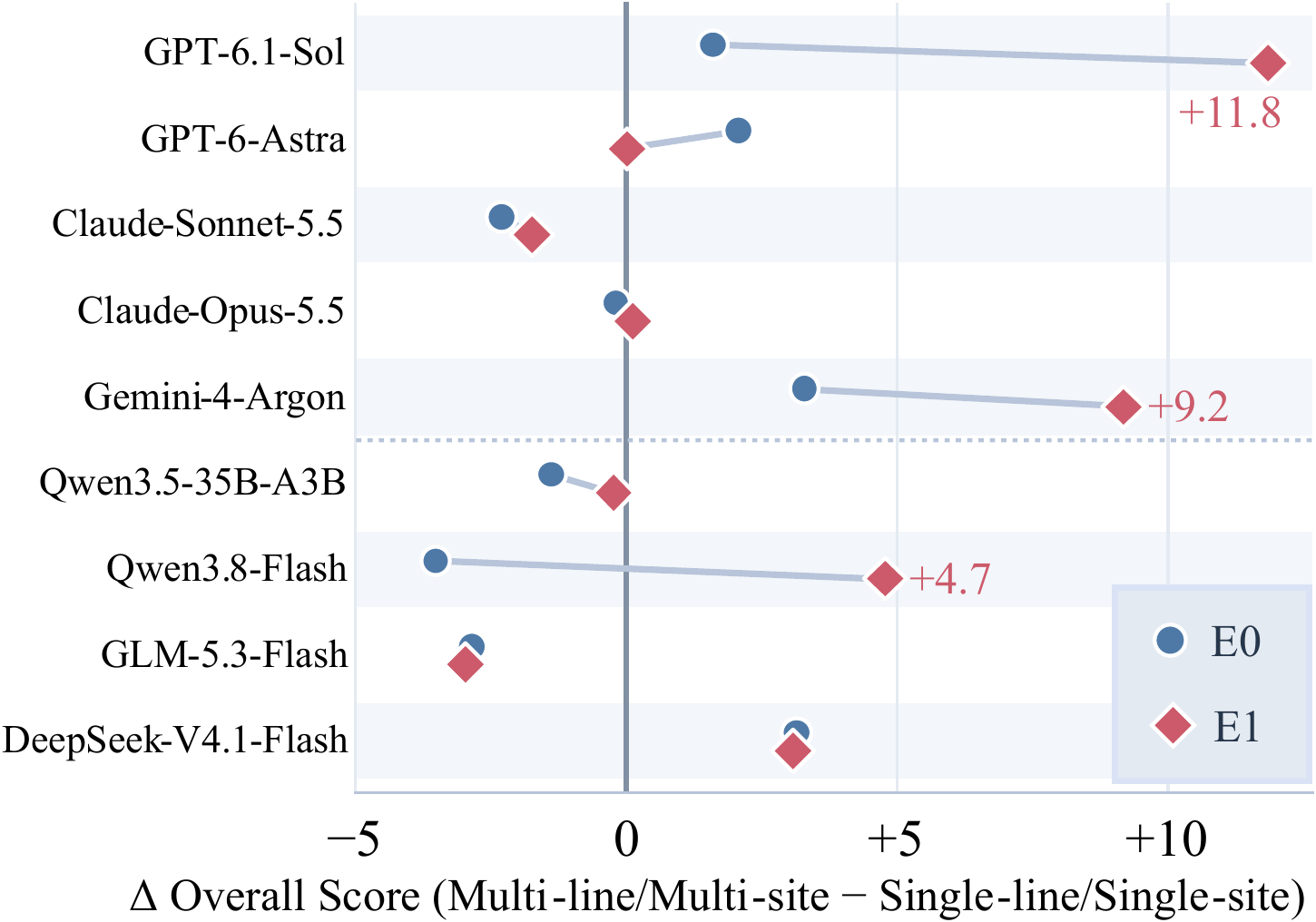}
\caption{Overall score differences by vulnerability patch scope \scriptsize under E0 and E1}
\label{fig4}
\vspace*{-\baselineskip}
\end{figure}

\subsection{Unified Evaluation Results}
Table~\ref{tab:unified-evaluation-results} presents the unified evaluation results of nine LLMs on SecRTL-Bench. Overall, model performance varies considerably, with the highest overall scores reaching 52.38 under E0 and 59.83 under E1, indicating substantial room for improvement in real-world IP-level RTL security analysis. Based on the results across models and input settings, we draw the following key findings:

\textbf{Limited transfer of security knowledge enhancement to code-level reasoning.} Compared with E0, E1 improves CWE MRR@5 for all nine models, with an average increase of 18.74 percentage points. However, only four models show improvements in localization F1 and repair success rate. Localization F1 and repair success rate exhibit consistent directional changes for eight of the nine models, with Qwen3.5-35B-A3B being the exception. These results indicate that additional CWE knowledge significantly improves vulnerability type identification but does not consistently enhance localization or repair, while localization accuracy may be more closely related to repair effectiveness.

\textbf{Substantial differences in repair conversion efficiency after vulnerability detection.} We measure repair conversion efficiency as the ratio of repair success rate to vulnerability detection rate, reflecting the proportion of detected vulnerabilities that are successfully repaired. Under E1, repair conversion efficiency ranges from 28.3\% to 96.0\%. Claude-Opus-5.5 achieves the highest detection rate (90.93\%) but a repair conversion efficiency of only 28.3\%. In contrast, Gemini-4-Argon has a lower detection rate (73.89\%) but achieves a repair conversion efficiency of 96.0\%. These results show that vulnerability detection and repair capabilities are not equivalent, and detection rate alone cannot adequately capture a model's end-to-end security analysis capabilities.

\textbf{Security verification as a key constraint on complete repair.} Across both experimental settings, all models achieve higher functional pass rates than security pass rates, while the gap between security pass rate and repair success rate is at most 2.45 percentage points. This indicates that most patches passing security verification also preserve normal functionality, but passing functional verification does not guarantee that the vulnerability has been eliminated. Therefore, relying solely on functional testing may overestimate repair capabilities, highlighting the need for independent security verification.

\begin{table*}[t]
\vspace*{-\baselineskip}
\caption{Independent Tasks Evaluation Results on SecRTL-Bench (\%). Best in bold, second-best underlined.}
\label{tab:independent-tasks-evaluation}
\centering
\tiny
\setlength{\tabcolsep}{3pt}
\renewcommand{\arraystretch}{1.05}
\setlength{\aboverulesep}{0.5pt}
\setlength{\belowrulesep}{0.5pt}
\resizebox{0.97\textwidth}{!}{%
\begin{tabular}{l*{14}{c}}
\hline
\multirow{3}{*}{\textbf{Models}} & \multicolumn{4}{c}{\textbf{Detection}} & \multicolumn{9}{c}{\textbf{Localization}} & \multicolumn{1}{c}{\textbf{Repair}} \\
\cmidrule(lr){2-5}\cmidrule(lr){6-14}\cmidrule(lr){15-15}
 & \multicolumn{2}{c}{\textbf{Detection Rate}} & \multicolumn{2}{c}{\textbf{CWE Hit@5}} & \multicolumn{3}{c}{\textbf{Precision}} & \multicolumn{3}{c}{\textbf{Recall}} & \multicolumn{3}{c}{\textbf{Macro-F1}} & \textbf{Repair Success} \\
\cmidrule(lr){2-3}\cmidrule(lr){4-5}\cmidrule(lr){6-8}\cmidrule(lr){9-11}\cmidrule(lr){12-14}\cmidrule(lr){15-15}
 & D0 & D1 & D0 & D1 & L0 & L1 & L2 & L0 & L1 & L2 & L0 & L1 & L2 & R0 \\
\hline
\multicolumn{15}{c}{\textit{Closed-source Models}} \\
\hline
GPT-6.1-Sol & 67.07 & 69.25 & \underline{20.82} & 50.85 & \underline{55.82} & \underline{61.86} & \underline{61.47} & 85.89 & \underline{96.02} & \underline{94.13} & \underline{61.31} & \underline{68.04} & \underline{67.18} & 92.98 \\
GPT-6-Astra & 61.17 & 68.45 & 8.74 & 46.84 & \textbf{64.06} & \textbf{69.53} & \textbf{70.94} & \underline{89.39} & \textbf{96.30} & \textbf{96.40} & \textbf{68.84} & \textbf{74.55} & \textbf{75.55} & \underline{97.34} \\
Claude-Sonnet-5.5 & \underline{82.73} & \underline{83.62} & 11.19 & 31.05 & 19.43 & 27.43 & 27.81 & 51.62 & 76.58 & 77.24 & 23.76 & 32.86 & 33.33 & 75.37 \\
Claude-Opus-5.5 & \textbf{83.50} & \textbf{90.93} & 8.98 & 25.98 & 17.10 & 24.10 & 22.69 & 54.66 & 80.66 & 81.40 & 21.44 & 29.93 & 28.83 & 72.66 \\
Gemini-4-Argon & 78.69 & 73.89 & \textbf{24.21} & \textbf{66.50} & 26.88 & 36.36 & 34.95 & \textbf{89.40} & 87.39 & 89.81 & 33.39 & 44.00 & 42.96 & \textbf{97.82} \\
\hline
\multicolumn{15}{c}{\textit{Open-source Models}} \\
\hline
Qwen3.5-35B-A3B & 42.93 & 70.54 & 0.25 & 26.49 & 4.99 & 10.58 & 11.50 & 21.58 & 17.16 & 43.99 & 6.58 & 13.66 & 14.98 & 44.44 \\
Qwen3.8-Flash & 43.80 & 64.27 & 0.73 & 34.49 & 21.98 & 46.15 & 45.11 & 52.06 & 82.38 & 81.34 & 26.97 & 52.04 & 50.69 & 90.46 \\
GLM-5.3-Flash & 32.76 & 38.83 & 3.18 & 21.01 & 23.80 & 34.14 & 36.87 & 55.73 & 79.93 & 82.72 & 28.67 & 41.99 & 44.97 & 83.62 \\
DeepSeek-V4.1-Flash & 72.64 & 75.18 & 19.85 & \underline{55.47} & 45.87 & 58.62 & 57.36 & 75.92 & 92.85 & 91.87 & 50.11 & 63.97 & 63.03 & 95.88 \\
\hline
\end{tabular}%
}
\vspace*{-\baselineskip}
\end{table*}

\textbf{Vulnerability Scope Analysis.} Fig.~\ref{fig4}  compares models' overall ability to handle single-line/single-site (331) and multi-line/multi-site (82) vulnerabilities. Scores to the left of zero favor single-line/single-site cases, while those to the right favor multi-line/multi-site cases. Under E0, four models favor multi-line/multi-site cases and five favor single-line/single-site cases, with gaps within 3.6 points. Under E1, GPT-6.1-Sol and Gemini-4-Argon show larger advantages on multi-line/multi-site cases (+11.8 and +9.2 points), while Qwen3.8-Flash shifts from a negative to a positive gap. These results suggest that multi-line/multi-site vulnerabilities are not consistently more challenging, and CWE knowledge can alter models' relative performance across vulnerability scopes.

\begin{figure}[t]
\centering
\includegraphics[width=0.95\columnwidth]{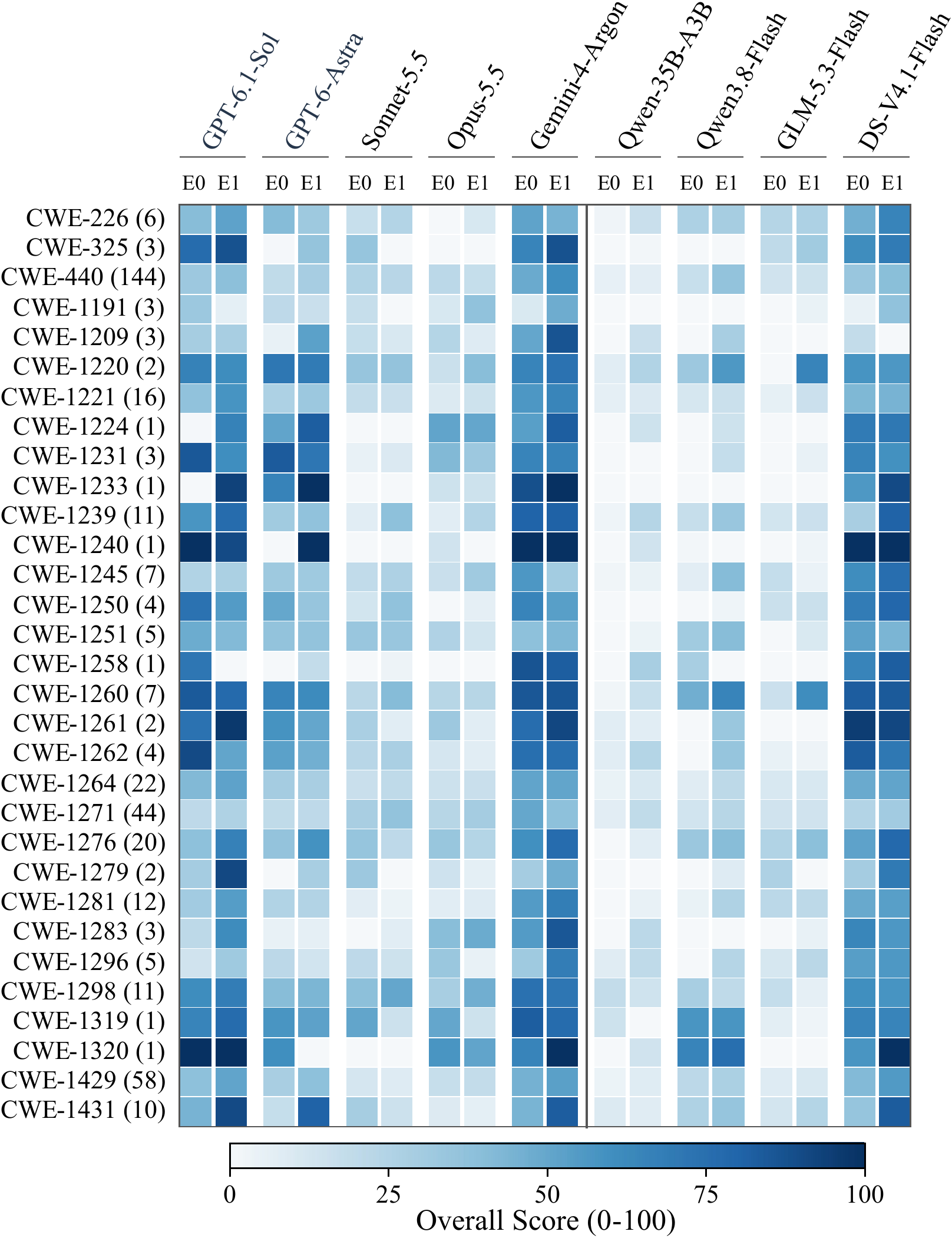}
\caption{CWE-wise overall scores across models under E0 and E1}
\label{fig5}
\vspace*{-\baselineskip}
\end{figure}

\textbf{CWE-wise Analysis.} Fig.~\ref{fig5} presents the overall scores of nine models across 31 Hardware CWE weaknesses under E0 and E1. The results show substantial performance variations across CWE categories, with no model consistently achieving high scores on all weaknesses. Moreover, the benefits of additional CWE knowledge differ across models and weakness types. These findings highlight the importance of CWE-level analysis for identifying model-specific strengths and limitations beyond aggregate evaluation scores.

\subsection{Independent Evaluation Results}
To assess task-specific capabilities, we evaluate detection, localization, and repair independently. For detection, D0 includes IP-level designs, build configurations, and normal functionality descriptions; D1 adds retrievable CWE knowledge base. For localization, L0 includes IP-level designs, build configurations, normal functionality descriptions, and a vulnerability-presence hint without revealing its type; L1 adds security requirements, and L2 adds the correct CWE description. For repair, R0 includes IP-level designs, build configurations, functional/security requirements, the correct CWE, and ground-truth vulnerability locations.

As shown in Table~\ref{tab:independent-tasks-evaluation}, D1 improves CWE hit@5 accuracy for all nine models, but does not consistently improve detection rates. For localization, L1 increases F1 across all nine models by an average of 11.1 points, whereas L2 provides almost no additional improvement (0.05 points). Under R0, six models achieve repair success rates above 80\%, led by Gemini-4-Argon (97.82\%) and GPT-6-Astra (97.34\%). Notably, Qwen3.8-Flash achieves 90.46\% repair success despite an L2 localization F1 of only 50.69\%. These results highlight vulnerability identification and localization as key bottlenecks in end-to-end RTL security repair, despite strong repair performance with ground-truth vulnerability information.

\section{Conclusion}
This paper presents SecRTL-Bench, a security benchmark based on real-world IP-level RTL designs for unified evaluation of LLMs across vulnerability detection, localization, and repair tasks. SecRTL-Bench employs multi-agent collaboration and security contracts to guide vulnerability construction, while combining automatically calibrated verification environments with EDA-based dynamic simulation to validate injected vulnerabilities. Evaluation results reveal the limitations of current LLMs in real-world RTL security analysis, particularly the substantial gap between vulnerability diagnosis and successful repair. Current models still struggle to translate identified security weaknesses into effective code patches that satisfy both functional and security requirements.

\clearpage
\bibliographystyle{IEEEtran}
\bibliography{references}

\end{document}